\documentclass[iicol,pdflatex,sn-mathphys-num]{sn-jnl}

\usepackage{amsmath}
\usepackage{flushend}
\usepackage{graphicx}

\hypersetup{colorlinks,citecolor=blue,urlcolor=blue,linkcolor=blue}

\begin{document}

\title{Study on net-proton number fluctuations in Au+Au collisions at $\sqrt{\textit{s}_{\text{NN}}}$ = 7.7--27 GeV with the SMASH model}

\author[1]{\fnm{Fengxi}\sur{Yan}}\equalcont{These authors contributed equally to this work.}
\author[1]{\fnm{Ting}\sur{Liu}}\equalcont{These authors contributed equally to this work.}
\author[2]{\fnm{Rongrong}\sur{Ma}}
\author[1]{\fnm{Yunshuang}\sur{Qin}}
\author[1]{\fnm{Shule}\sur{Wu}}
\author*[1]{\fnm{Shenghui}\sur{Zhang}}\email{zhangshh08@cqu.edu.cn}
\author*[3]{\fnm{Fan}\sur{Si}}\email{fsi@physi.uni-heidelberg.de}

\affil[1]{Department of Physics and Chongqing Key Laboratory for Strongly Coupled Physics, Chongqing University, Chongqing 401331, People's Republic of China}
\affil[2]{Physics Department, Brookhaven National Laboratory, Upton, New York 11973, USA}
\affil[3]{Physikalisches Institut, Ruprecht-Karls-Universit\"{a}t Heidelberg, 69120 Heidelberg, Germany}
\date{\today}

\abstract{
    Higher-order cumulants ($C_n$) of the event-by-event net-proton number ($\Delta N_{\mathrm{p}} = N_{\mathrm{p}} - N_{\bar{\mathrm{p}}}$) distributions are sensitive observables in the search for the QCD critical point. A reliable quantitative baseline for non-critical effects is essential for interpreting these observables. We calculate $C_2/\langle N_{\mathrm{p}} + N_{\bar{\mathrm{p}}}\rangle$, $C_3/C_1$, and $C_4/C_2$ in Au+Au collisions at collision energies $\sqrt{s_{\mathrm{NN}}} = 7.7$--$27$~GeV with the SMASH hadronic transport model. The collision-centrality, rapidity-window, and collision-energy dependences are reported. The cumulant ratios are predicted below the independent-Poisson baseline, with the suppression increasing toward more central collisions and larger rapidity windows. In 70--80\% collisions, SMASH calculations lie above the STAR measurements for all three ratios, whereas in 0--5\% collisions, the STAR $C_4/C_2$ value at 19.6~GeV lies below the SMASH reference with a pull of approximately $-2.7$. These results provide an independent non-critical hadronic baseline for assessing the model dependence of the observed data–model discrepancies.
}

\maketitle

\section{Introduction}

The strong interaction of quarks and gluons is described by quantum chromodynamics (QCD)~\cite{Fritzsch:1973pi}. The phase structure of strongly interacting matter is conventionally represented by the QCD phase diagram in the temperature ($T$)--baryon chemical potential ($\mu_{\mathrm{B}}$) plane~\cite{Fukushima:2010bq}. Lattice-QCD calculations establish a smooth crossover between hadronic matter and the quark--gluon plasma at low
$\mu_{\mathrm{B}}$~\cite{Aoki:2006we,HotQCD:2018pds}. At larger $\mu_{\mathrm{B}}$, a first-order phase-transition line may exist and terminate at the potential QCD critical point~\cite{Fukushima:2010bq}. In relativistic heavy-ion collisions, varying the collision energy ($\sqrt{s_{\mathrm{NN}}}$) changes the thermodynamic region accessed by the produced nuclear matter, with different chemical freeze-out values of $T$ and $\mu_{\mathrm{B}}$~\cite{Andronic:2026cdu}.

Near the QCD critical point, the correlation length ($\xi$) is expected to increase. Fluctuations of conserved charges ($q$), such as net electric charge, net-baryon number, and net strangeness, are sensitive to $\xi$~\cite{Luo:2017faz} and directly related to the corresponding $n$th-order QCD susceptibility ($\chi_n^q$)~\cite{Friman:2011pf}. In the grand-canonical ensemble, the $n$th-order cumulant of the conserved charge and susceptibility are related by $C_n^q = VT^3\chi_n^q$, where $V$ is the system volume. Ratios of cumulants reduce the leading volume dependence and can be compared with the corresponding susceptibility ratios~\cite{Asakawa:2009aj,Friman:2011pf}. Critical contributions grow rapidly with $\xi$ and may generate characteristic magnitudes in higher-order cumulants and their ratios~\cite{Stephanov:2008qz,Athanasiou:2010kw,Stephanov:2011pb,Bzdak:2019pkr}. Experimentally, the cumulants of the net-proton number can be used as proxies for those of the experimentally inaccessible net-baryon number~\cite{Hatta:2003wn,Kitazawa:2011wh}. In particular, the net-proton cumulant ratio $C_4/C_2$ is predicted to exhibit a non-monotonic collision-energy dependence relative to a non-critical baseline near the QCD critical point~\cite{Stephanov:2011pb}.

The same observables are also affected by non-critical physics. Finite acceptance, exact baryon-number conservation, resonance decays, hadronic rescattering, participant and volume fluctuations, and the centrality estimator can all modify measured cumulants~\cite{Bzdak:2012ab,Bzdak:2012an,Skokov:2012ds,Luo:2017faz,Vovchenko:2020gne}. These effects depend on collision energy, centrality, and phase-space acceptance and therefore cannot in general be represented by a single energy-independent reference.

The Beam Energy Scan (BES) program at the Relativistic Heavy Ion Collider (RHIC) was established to map the QCD phase diagram by conducting Au+Au collisions at $\sqrt{s_{\mathrm{NN}}} = 3$--$200$~GeV~\cite{STAR:2010vob,Bzdak:2019pkr}. The most recent measurement at $7.7$--$27$~GeV by the STAR Collaboration exhibits a minimum of $C_4/C_2$ around 19.6~GeV in central collisions, with deviations of approximately $2$--$5\sigma$ from non-critical references~\cite{STAR:2025zdq}, depending on the model without a critical point or the peripheral-collision data chosen as the reference. In order to accurately understand the potential critical signals from the experimental data, it is crucial to establish more comparable and reliable baselines without any critical elements.

Hadronic transport models provide non-critical baselines by incorporating microscopic scattering, resonance formation and decay, conservation laws, and the dynamical evolution of the hadronic system without introducing critical fluctuations. The Ultra-relativistic Quantum Molecular Dynamics (UrQMD) model has been widely used in fluctuation studies~\cite{Bass:1998ca,Bleicher:1999xi}. Here, we construct an independent baseline with the Simulating Many Accelerated Strongly-interacting Hadrons (SMASH) model, which solves the relativistic Boltzmann equation through the propagation and interaction of individual hadrons~\cite{SMASH:2016zqf}. Its microscopic evolution includes binary scattering and resonance formation and decay, while string excitation and fragmentation extend the description to the collision energies considered here~\cite{Mohs:2019iee}. The separately developed transport and particle-production implementations in the SMASH model complement UrQMD and allow the model dependence of a hadronic baseline to be assessed.

In this article, we study the collision-centrality, rapidity-window, and collision-energy dependences of cumulant ratios with the SMASH model and compare the results with STAR and UrQMD. This provides an independent test of which data--model discrepancies persist across transport implementations and how their magnitudes depend on the reference. Certainly, the SMASH model comparison benefits from matching the experimental (anti)proton acceptance and collision-centrality determination.

\section{SMASH event sample and observables}

\begin{figure*}
    \centering
    \includegraphics[width=\linewidth]{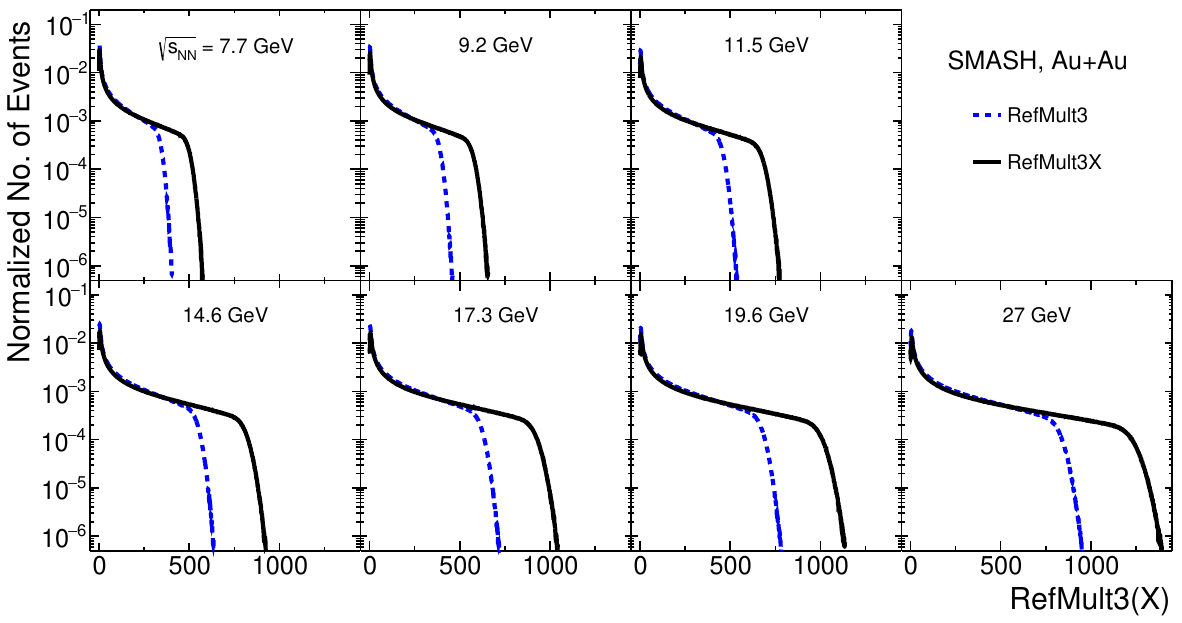}
    \caption{Normalized RefMult3 (dashed blue curve) and RefMult3X (solid black curve) distributions in SMASH Au+Au collisions at $\sqrt{s_{\mathrm{NN}}} = 7.7$--$27$~GeV.}
    \label{fig:reference_multiplicity}
\end{figure*}

We generate Au+Au collision events with SMASH version 3.3~\cite{SMASH:3.3} at seven energies: $\sqrt{s_{\mathrm{NN}}} = 7.7$, $9.2$, $11.5$, $14.6$, $17.3$, $19.6$, and $27$~GeV. Approximately $3 \times 10^7$ events are analyzed at each energy from 7.7 to 11.5~GeV, and $6 \times 10^7$ at each energy from 14.6 to 27~GeV. The energy, momentum, net electric charge, net-baryon number, and net strangeness are conserved event by event. No critical dynamics or fluctuations are introduced. The sample therefore retains conventional effects such as conservation laws, resonance decays, scattering, baryon stopping, and particle production while serving as a non-critical reference.

For each event, the proton and antiproton numbers inside the selected acceptance are denoted by $N_{\mathrm{p}}$ and $N_{\bar{\mathrm{p}}}$, respectively. The net-proton number is given by $\Delta N_{\mathrm{p}} = N_{\mathrm{p}} - N_{\bar{\mathrm{p}}}$. Using $\delta\Delta N_{\mathrm{p}} = \Delta N_{\mathrm{p}} - \langle\Delta N_{\mathrm{p}}\rangle$, the first four cumulants are defined as
\begin{align}
    C_1 &= \langle\Delta N_{\mathrm{p}}\rangle,\\
    C_2 &= \langle(\delta\Delta N_{\mathrm{p}})^2\rangle,\\
    C_3 &= \langle(\delta\Delta N_{\mathrm{p}})^3\rangle,\\
    C_4 &= \langle(\delta\Delta N_{\mathrm{p}})^4\rangle - 3\langle(\delta\Delta N_{\mathrm{p}})^2\rangle^2.
\end{align}
The observables studied below are $C_2/\langle N_{\mathrm{p}} + N_{\bar{\mathrm{p}}}\rangle$, $C_3/C_1$, and $C_4/C_2$. For independent Poisson proton and antiproton numbers, all three equal unity. Their deviations from unity could reflect physical effects such as the collision dynamics and baryon conservation.

\section{Analysis method}

\subsection{Reference multiplicity and collision centrality}

The collision centrality is determined by the reference charged-particle multiplicity following the STAR analysis~\cite{STAR:2025zdq}. To avoid self-correlations, protons and antiprotons are excluded from the reference multiplicity~\cite{Luo:2013bmi}. Two reference multiplicities counting $\pi^{\pm}$ and $\mathrm{K}^{\pm}$ mesons within different pseudorapidity ($\eta$) ranges are defined: RefMult3 ($|\eta| < 1.0$) and RefMult3X ($|\eta| < 1.6$). The multiplicity distributions are shown in Fig.~\ref{fig:reference_multiplicity}. Each reference multiplicity distribution at each energy is integrated from high to low multiplicity. This procedure determines the 0--5\% and 5--10\% classes, followed by 10\%-wide classes through 70--80\%.

\begin{figure}
    \centering
    \includegraphics[width=\linewidth]{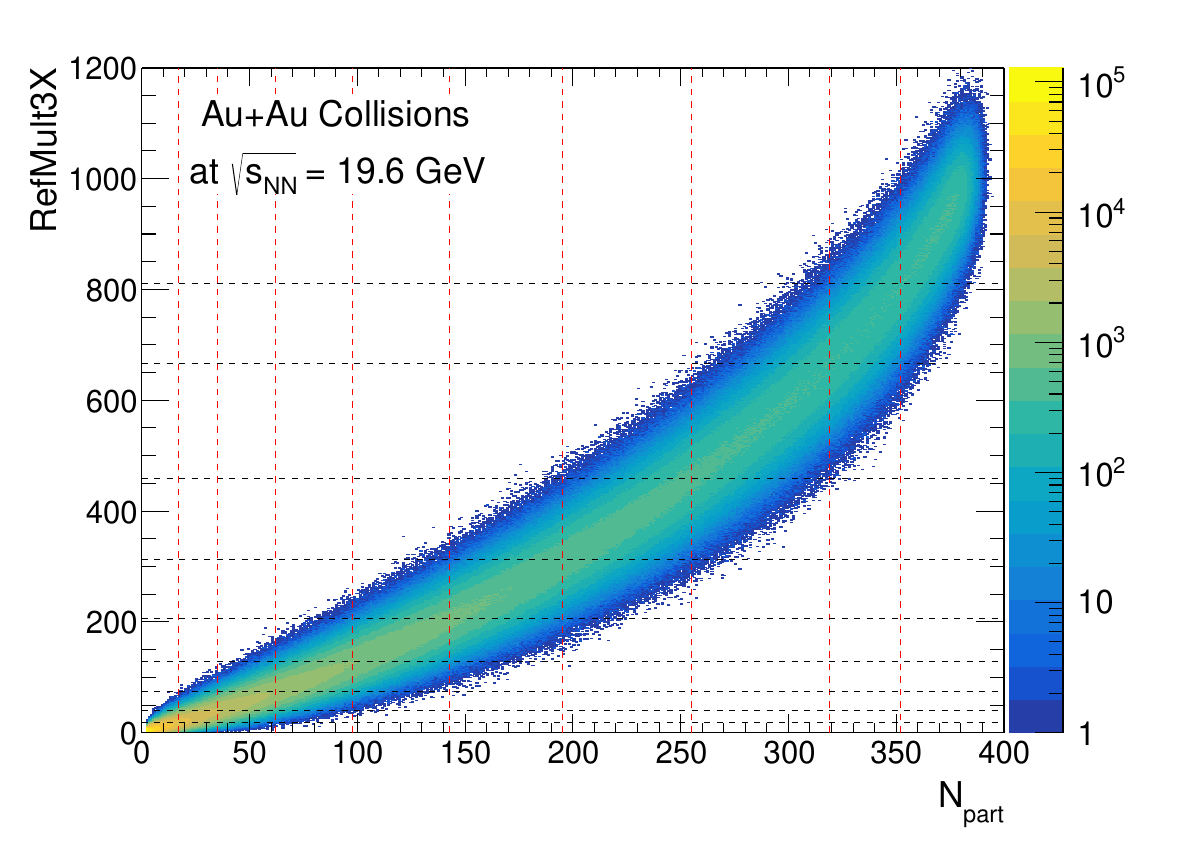}
    \caption{Correlation distribution of RefMult3X versus $N_{\mathrm{part}}$ in SMASH Au+Au collisions at $\sqrt{s_{\mathrm{NN}}} = 19.6$~GeV. Vertical and horizontal dashed lines indicate centrality-percentile boundaries selected by $N_{\mathrm{part}}$ and RefMult3X, respectively.}
    \label{fig:npart_refmult}
\end{figure}

Figure~\ref{fig:npart_refmult} shows the event-by-event correlation distribution of RefMult3X and the number of participating nucleons ($N_{\mathrm{part}}$) in 19.6 GeV Au+Au collisions as an example. Here, $N_{\mathrm{part}}$ is calculated as the difference between the initial nucleon number and the final spectator number in each event. A strong correlation is observed, but the relation is not one-to-one and shows a finite spread, which reflects finite centrality resolution and contributes residual volume fluctuations to the cumulants. RefMult3X provides better centrality resolution than RefMult3 due to larger multiplicity with larger $\eta$ acceptance~\cite{Luo:2013bmi,Chatterjee:2019fey}.

\subsection{Centrality-bin-width correction}

To suppress the volume fluctuation measured in the net-proton number fluctuations in each broad centrality class, the centrality-bin-width correction (CBWC)~\cite{Luo:2013bmi} is applied. The net-proton number cumulants are calculated in each unit-multiplicity bin, rather than in a broad centrality class. The corrected cumulants in a centrality class are obtained as the averaged cumulants in multiplicity bins:
\begin{equation}
    C_n^{\mathrm{CBWC}} = \frac{\sum_i N_{\mathrm{event},i} C_{n,i}}{\sum_i N_{\mathrm{event},i}},
\end{equation}
where the multiplicity index $i$ runs within the centrality class of interest. $N_{\mathrm{event},i}$ and $C_{n,i}$ represent the number of events and the $n$th-order cumulant, respectively, in the $i$th multiplicity bin.

\begin{figure}
    \centering
    \includegraphics[width=\linewidth]{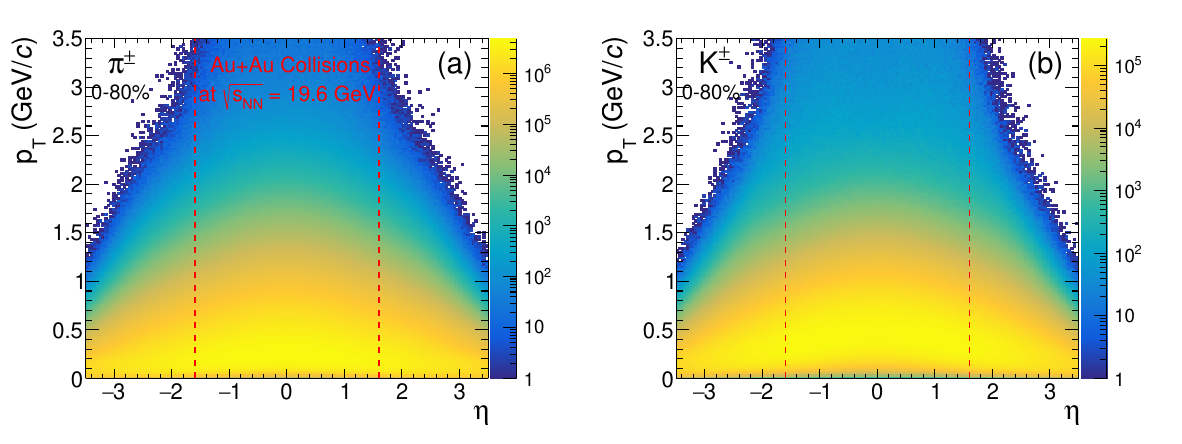}
    \caption{Distributions of transverse momentum ($p_{\mathrm{T}}$) versus pseudorapidity ($\eta$) of $\pi^{\pm}$ (a) and $\mathrm{K}^{\pm}$ (b) in 0--80\% Au+Au collisions at $\sqrt{s_{\mathrm{NN}}} = 19.6$~GeV from the SMASH model. The red dashed lines indicate selection cuts by $|\eta| < 1.6$, the RefMult3X acceptance.}
    \label{fig:meson}
\end{figure}

\subsection{Statistical uncertainties}

Statistical uncertainties of cumulants are calculated using the delta theorem approach~\cite{Luo:2011tp,Si:2021vog}. Raw moments up to twice the maximum cumulant order are used to construct the covariance matrix of the cumulants. The error formula for a generic cumulant ratio is
\begin{equation}
    \sigma_{C_m/C_n}^2 = \frac{\sigma_{C_m}^2}{C_n^2} + \frac{C_m^2\sigma_{C_n}^2}{C_n^4} - \frac{2C_m\operatorname{Cov}(C_m,C_n)}{C_n^3}.
\end{equation}
The cumulant errors from unit-multiplicity bins are combined to obtain the errors of CBWC-corrected cumulants as
\begin{equation}
    \sigma_{C_n^{\mathrm{CBWC}}} = \frac{\sqrt{\sum_i N_{\mathrm{event},i}^2\sigma_{C_{n,i}}^2}}{\sum_i N_{\mathrm{event},i}}.
\end{equation}

\begin{figure}
    \centering
    \includegraphics[width=\linewidth]{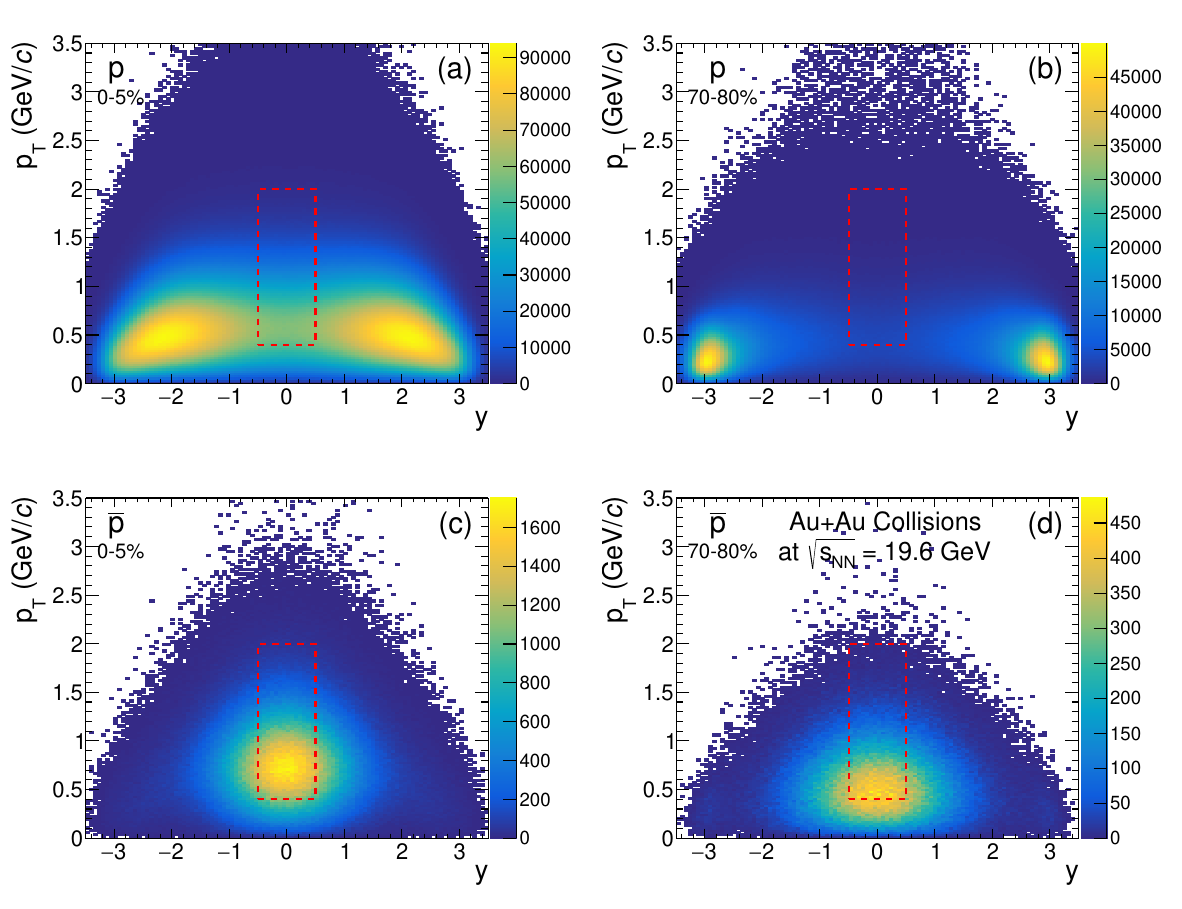}
    \caption{Distributions of transverse momentum ($p_{\mathrm{T}}$) versus rapidity ($y$) of protons (a) and antiprotons (c) in 0--5\% central Au+Au collisions at $\sqrt{s_{\mathrm{NN}}} = 19.6$~GeV from the SMASH model. Panels (b) and (d) show the corresponding distributions for the 70--80\% centrality. The red dashed boxes show the analysis acceptance, $0.4 < p_{\mathrm{T}} < 2.0$~GeV/$c$ and $|y| < 0.5$.}
    \label{fig:baryon}
\end{figure}

\begin{figure*}
    \centering
    \includegraphics[width=0.8\linewidth]{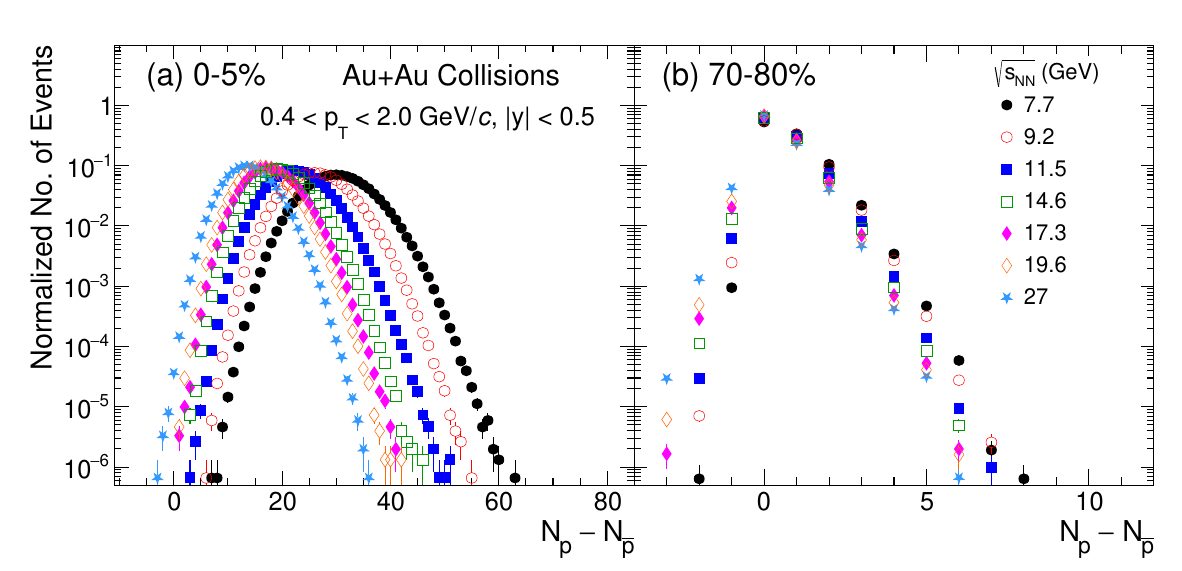}
    \caption{Normalized event-by-event distributions of the net-proton number ($\Delta N_{\mathrm{p}}$) in (a) central 0--5\% and (b) peripheral 70--80\% Au+Au collisions at $\sqrt{s_{\mathrm{NN}}} = 7.7$--$27$~GeV from the SMASH model. Protons and antiprotons satisfy $0.4 < p_{\mathrm{T}} < 2.0$~GeV/$c$ and $|y| < 0.5$.}
    \label{fig:netp_dist}
\end{figure*}

\begin{figure*}
    \centering
    \includegraphics[width=\linewidth]{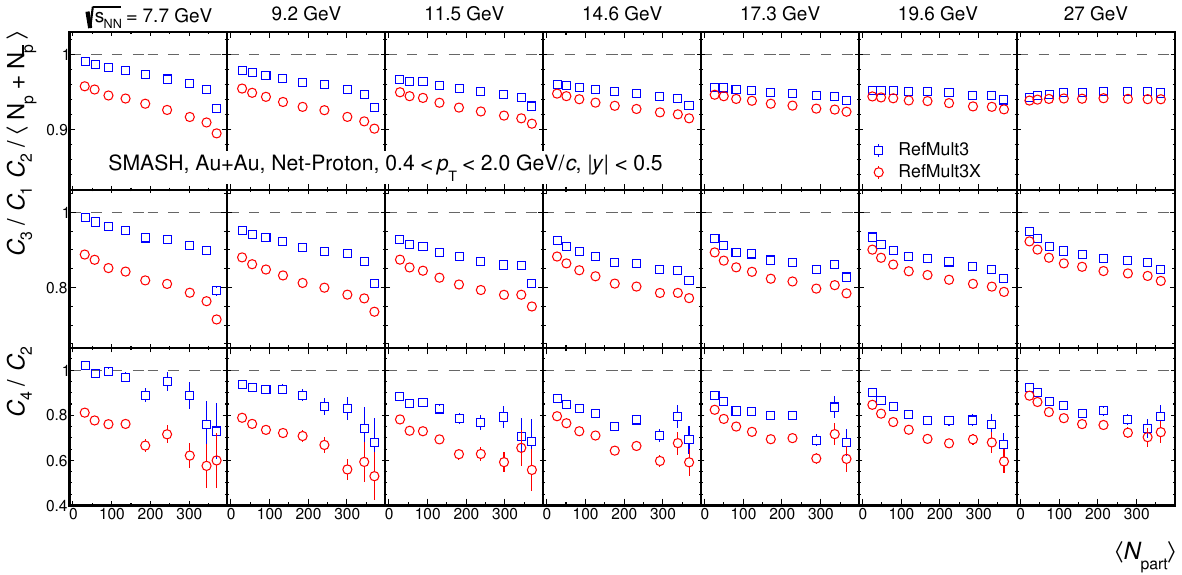}
    \caption{Collision-centrality (denoted as $\langle N_{\mathrm{part}}\rangle$) dependence of net-proton number cumulant ratios, $C_2/\langle N_{\mathrm{p}} + N_{\bar{\mathrm{p}}}\rangle$ (top panels), $C_3/C_1$ (middle panels), and $C_4/C_2$ (bottom panels), in SMASH Au+Au collisions at $\sqrt{s_{\mathrm{NN}}} = 7.7$--$27$~GeV. Blue squares and red circles denote results obtained using RefMult3 and RefMult3X for the centrality determination, respectively. The acceptance is $0.4 < p_{\mathrm{T}} < 2.0$~GeV/$c$ and $|y| < 0.5$. Dashed horizontal lines mark unity as the Poisson baseline.}
    \label{fig:ratios_centrality}
\end{figure*}

\subsection{Particle acceptance}

Figure~\ref{fig:meson} shows the phase-space distributions (transverse momentum ($p_{\mathrm{T}}$) versus $\eta$) of the charged mesons, $\pi^{\pm}$ in panel (a) and $\mathrm{K}^{\pm}$ in panel (b), used for centrality determination in 0--80\% Au+Au collisions at $\sqrt{s_{\mathrm{NN}}} = 19.6$~GeV. The broad $\eta$ distribution provides a reference sample separated by particle species from the proton and antiproton observables. The red dashed lines show the RefMult3X $\eta$ interval.

The phase-space distributions ($p_{\mathrm{T}}$ versus rapidity ($y$)) of the protons and antiprotons, which determine the particle numbers used for the fluctuation analysis, in Au+Au collisions at $\sqrt{s_{\mathrm{NN}}} = 19.6$~GeV are shown in Fig.~\ref{fig:baryon}. Upper (lower) panels display protons (antiprotons), and left (right) panels present central (peripheral) events. The proton distributions are seen to peak in the forward and backward rapidity regions, which are associated with transported baryons, while antiprotons accumulate around midrapidity. The numbers of both in central collisions are larger than those in the peripheral collisions. The kinematic acceptance of the analysis, $0.4 < p_{\mathrm{T}} < 2.0$~GeV/$c$ and $|y| < 0.5$, is indicated by the red dashed box in each panel, and the upper cut of $|y|$ ($|y|_{\mathrm{max}}$) is varied to 0.1, 0.2, 0.3, 0.4, and 0.6 when studying the rapidity-window dependence.

\section{Results}

\begin{figure*}
    \centering
    \includegraphics[width=\linewidth]{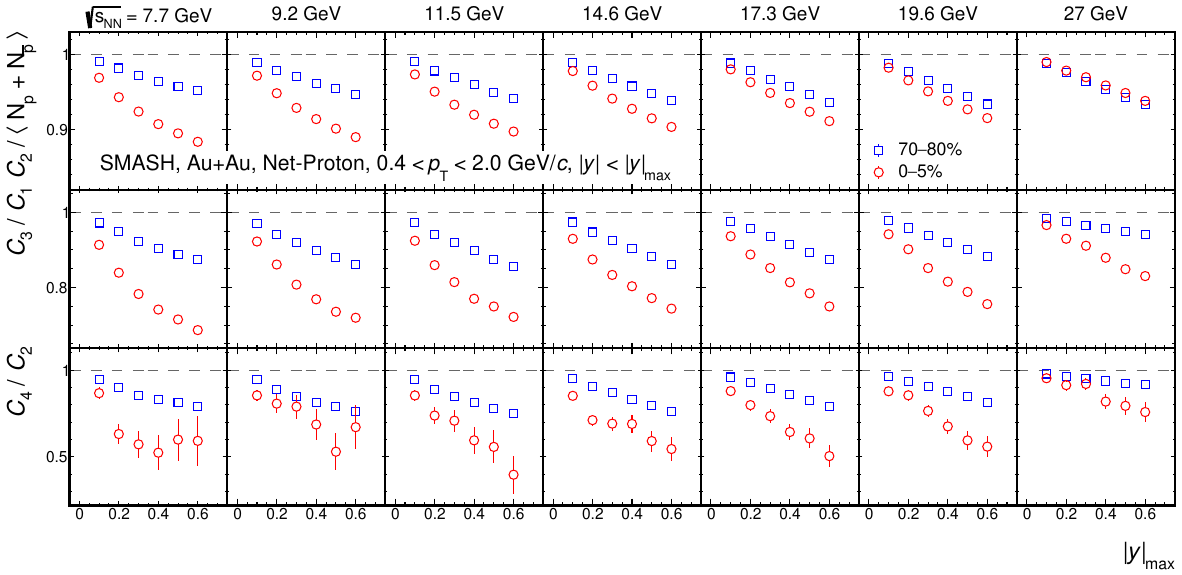}
    \caption{Rapidity-window dependence of net-proton number cumulant ratios, $C_2/\langle N_{\mathrm{p}} + N_{\bar{\mathrm{p}}}\rangle$ (top panels), $C_3/C_1$ (middle panels), and $C_4/C_2$ (bottom panels), in SMASH Au+Au collisions at $\sqrt{s_{\mathrm{NN}}} = 7.7$--$27$~GeV. Blue squares and red circles denote 70--80\% peripheral and 0--5\% central collisions, respectively. RefMult3X is used from 7.7 to 19.6~GeV and RefMult3 for 27~GeV. The $p_{\mathrm{T}}$ interval is $0.4 < p_{\mathrm{T}} < 2.0$~GeV/$c$. Dashed horizontal lines mark unity as the Poisson baseline.}
    \label{fig:ratios_rapidity}
\end{figure*}

\subsection{Net-proton number distributions}

The distributions of net-proton numbers within $0.4 < p_{\mathrm{T}} < 2.0$~GeV/$c$ and $|y| < 0.5$ in SMASH Au+Au collisions at $\sqrt{s_{\mathrm{NN}}} = 7.7$--$27$~GeV are shown in Fig.~\ref{fig:netp_dist} and provide the input to the cumulants. The mean moves to larger positive $\Delta N_{\mathrm{p}}$ for lower collision energy, for both the 0--5\% centrality in panel (a) and the 70--80\% centrality in panel (b). This phenomenon is more pronounced in more central collisions, reflecting stronger baryon stopping and reduced relative contribution from proton--antiproton pairs. The distributions are broader in 0--5\% central collisions due to the larger accepted proton and antiproton numbers. In peripheral collisions, the distributions are narrower and concentrated near zero.

\begin{figure*}
    \centering
    \includegraphics[width=\linewidth]{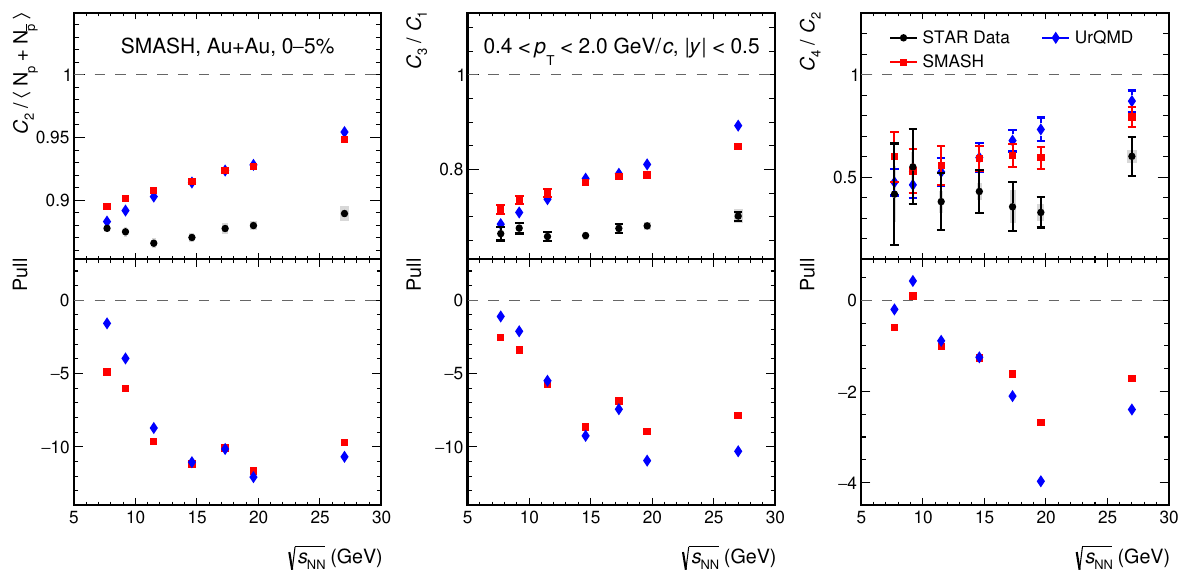}
    \caption{Collision-energy dependence of the net-proton number cumulant ratios, $C_2/\langle N_{\mathrm{p}} + N_{\bar{\mathrm{p}}}\rangle$ (left panels), $C_3/C_1$ (middle panels), and $C_4/C_2$ (right panels), in 0--5\% central Au+Au collisions. The upper panels compare STAR data (black circles)~\cite{STAR:2025zdq}, UrQMD calculations (blue diamonds)~\cite{STAR:2025zdq}, and SMASH ones (red squares). Vertical bars denote statistical uncertainties, and gray boxes of the STAR data display systematic uncertainties. The lower panels show the data--model pull defined in Eq.~\eqref{equ:pull}. Dashed lines mark unity as the Poisson baseline in the upper panels and zero in the lower panels.}
    \label{fig:ratios05}
\end{figure*}

\subsection{Collision-centrality dependence}

The collision-centrality (denoted as $\langle N_{\mathrm{part}}\rangle$) dependence of all net-proton number cumulant ratios, $C_2/\langle N_{\mathrm{p}} + N_{\bar{\mathrm{p}}}\rangle$, $C_3/C_1$, and $C_4/C_2$ at different energies is shown in Fig.~\ref{fig:ratios_centrality}. Both results from the centrality determination using RefMult3 and RefMult3X are presented. The cumulant ratios generally decrease with increasing $\langle N_{\mathrm{part}}\rangle$, a suppression qualitatively consistent with the effects of baryon-number conservation in a finite acceptance~\cite{Bzdak:2012an}. The trends are smooth for $C_2/\langle N_{\mathrm{p}} + N_{\bar{\mathrm{p}}}\rangle$ and $C_3/C_1$, whereas $C_4/C_2$ shows larger uncertainties and visible point-to-point fluctuations, especially at the lowest energies.

The cumulant ratios from RefMult3X are generally smaller than those from RefMult3 at the same collision energy and centrality. The larger $\eta$ coverage of RefMult3X improves the centrality resolution and reduces the effect caused by residual volume variation. The difference is generally larger at lower collision energies, indicating the importance of the reference-multiplicity choice.

\subsection{Rapidity-window dependence}

Figure~\ref{fig:ratios_rapidity} shows the dependence of net-proton number cumulant ratios, $C_2/\langle N_{\mathrm{p}} + N_{\bar{\mathrm{p}}}\rangle$, $C_3/C_1$, and $C_4/C_2$, on the rapidity-window varying through the symmetric intervals $|y| < 0.1$, $0.2$, $0.3$, $0.4$, $0.5$, and $0.6$ at different energies. Centrality determination uses RefMult3X for $\sqrt{s_{\mathrm{NN}}} = 7.7$--$19.6$~GeV and RefMult3 for 27~GeV, as in the STAR analysis~\cite{STAR:2025zdq} and in the following study of collision-energy dependence. The cumulant ratios approach unity as the acceptance becomes narrow and generally decrease when the window is enlarged. The suppression is generally stronger in 0--5\% central than in 70--80\% peripheral collisions. This behavior is consistent with global baryon-number conservation: enlarging the acceptance samples a larger fraction of the conserved charge and strengthens the anti-correlations between the accepted and unaccepted regions~\cite{Bzdak:2012an}.

\subsection{Collision-energy dependence}

In the upper panels of Fig.~\ref{fig:ratios05}, the collision-energy dependences of net-proton number cumulant ratios, $C_2/\langle N_{\mathrm{p}} + N_{\bar{\mathrm{p}}}\rangle$, $C_3/C_1$, and $C_4/C_2$, are shown and compared to those of the STAR data and UrQMD model in the common acceptance $0.4 < p_{\mathrm{T}} < 2.0$~GeV/$c$ and $|y| < 0.5$ for 0--5\% central collisions. Both STAR and UrQMD results are quoted from Ref.~\cite{STAR:2025zdq}. UrQMD and SMASH models yield broadly similar magnitudes and energy dependences for the cumulant ratios. The $C_2/\langle N_{\mathrm{p}} + N_{\bar{\mathrm{p}}}\rangle$ and $C_3/C_1$ values in the UrQMD and SMASH models increase monotonically from 7.7 to 27~GeV and lie systematically above the STAR data~\cite{STAR:2025zdq}. For $C_4/C_2$, both model references also rise overall with energy, while the STAR points exhibit a localized minimum near $\sqrt{s_{\mathrm{NN}}}=19.6$~GeV and a recovery at 27~GeV. SMASH calculations lie below UrQMD ones from 17.3 to 27~GeV and are consequently closer to the STAR $C_4/C_2$ values in this interval.

The lower panels of Fig.~\ref{fig:ratios05} quantify the signed data--model difference given by
\begin{equation}
    \mathrm{Pull} = \frac{Y_{\mathrm{data}} - Y_{\mathrm{model}}}{\sqrt{\sigma_{\mathrm{data,stat}}^2 + \sigma_{\mathrm{data,sys}}^2 + \sigma_{\mathrm{model,stat}}^2}},
    \label{equ:pull}
\end{equation}
where $Y$ denotes the cumulant ratio. All three uncertainty components are treated as independent in this diagnostic quantity. A negative pull means that the STAR value is below the model reference. The $C_2/\langle N_{\mathrm{p}} + N_{\bar{\mathrm{p}}}\rangle$ and $C_3/C_1$ pulls are negative for each energy and reach large magnitudes because the differences are systematic compared with the small uncertainties. For $C_4/C_2$, the strongest discrepancy occurs at $\sqrt{s_{\mathrm{NN}}}=19.6$~GeV: the pulls relative to SMASH and UrQMD results are approximately $-2.7$ and $-4.0$, respectively. The difference between the two pulls illustrates the model dependence of the quoted deviation from a non-critical reference.

\begin{figure*}
    \centering
    \includegraphics[width=\linewidth]{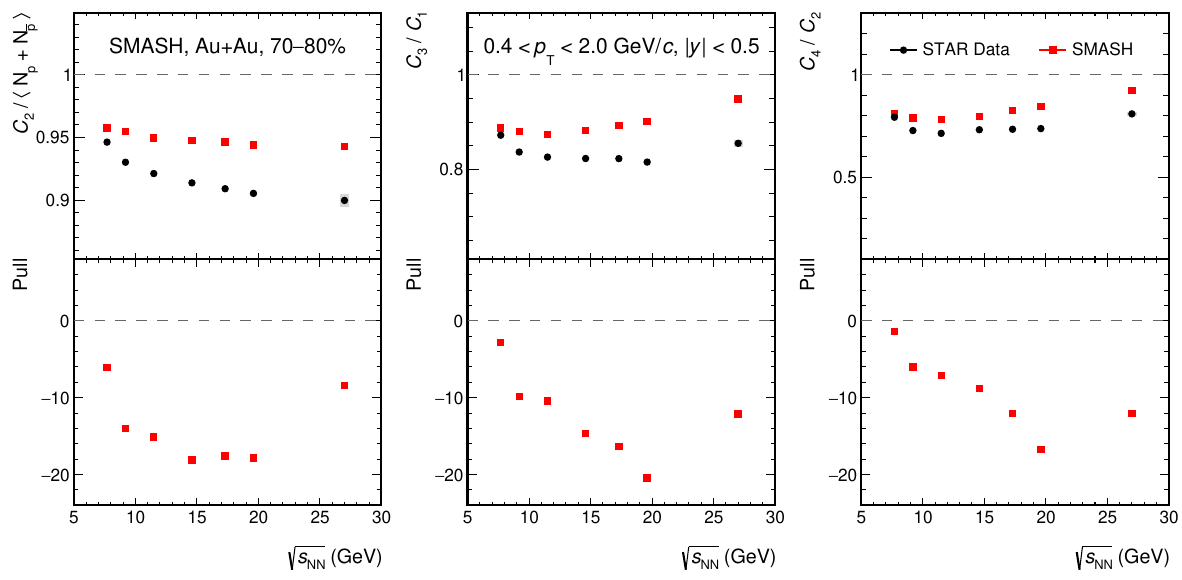}
    \caption{Collision-energy dependence of the net-proton cumulant ratios in 70--80\% peripheral Au+Au collisions. The upper panels compare STAR data (black circles) with SMASH results (red squares). Vertical bars denote statistical uncertainties, and gray boxes of the STAR data display systematic uncertainties. The lower panels show the data--model pull defined in Eq.~\eqref{equ:pull}. Dashed lines mark unity as the Poisson baseline in the upper panels and zero in the lower panels.}
    \label{fig:ratios7080}
\end{figure*}

Figure~\ref{fig:ratios7080} presents the corresponding comparison between STAR and SMASH results for 70--80\% peripheral collisions. Their cumulant ratios have qualitatively similar energy dependence: $C_2/\langle N_{\mathrm{p}} + N_{\bar{\mathrm{p}}}\rangle$ decreases with increasing energy, while $C_3/C_1$ and $C_4/C_2$ display shallow minima in the intermediate-energy region. Quantitatively, however, SMASH calculations lie above STAR measurements for all three observables and all seven energies, with correspondingly sizable negative pulls throughout the energy range.

Taken together, the two energy-dependent comparisons show that the SMASH model supplies an effective non-critical baseline but does not reproduce the absolute STAR values. In central collisions, SMASH and UrQMD models provide generally similar results, lending support to the main features of the hadronic transport baseline. Both models fail to reproduce the localized minimum in the STAR $C_4/C_2$ measurement near $\sqrt{s_{\mathrm{NN}}}=19.6$~GeV, although the SMASH calculation lies closer to the experimental data. In peripheral collisions, the systematic offsets across all energies reveal limitations of the present baseline in quantitatively describing the fluctuations.

\section{Conclusion}

Net-proton number fluctuations in Au+Au collisions at $\sqrt{s_{\mathrm{NN}}} = 7.7$--$27$~GeV have been studied with the SMASH hadronic transport model. The analysis implements the STAR (anti)proton acceptance and the energy-dependent RefMult3 or RefMult3X centrality definitions. Cumulants are calculated in unit reference-multiplicity bins and corrected by CBWC, while statistical uncertainties are obtained analytically with the Delta theorem and include the relevant cumulant covariances.

The event-by-event net-proton number distributions broaden toward central collisions and shift to larger positive values as the energy decreases, reflecting the evolution of baryon stopping. All three cumulant ratios, $C_2/\langle N_{\mathrm{p}} + N_{\bar{\mathrm{p}}}\rangle$, $C_3/C_1$, and $C_4/C_2$, generally decrease as collisions become more central or as the rapidity acceptance is enlarged. The comparison between RefMult3 and RefMult3X shows that the centrality resolution remains important after CBWC, particularly for lower energies.

UrQMD and SMASH calculations provide smooth energy dependences in central collisions, whereas the STAR measurement of $C_4/C_2$ exhibits a localized minimum near $\sqrt{s_{\mathrm{NN}}} = 19.6$~GeV. With the uncertainty prescription used here, the corresponding STAR--SMASH and STAR--UrQMD pulls are approximately $-2.7$ and $-4.0$. In 70--80\% collisions, SMASH results follow the broad energy trends but systematically exceed all three ratios from STAR with the small uncertainties, leading to sizable negative pulls at all energies. These results provide an independent SMASH baseline complementary to the UrQMD one, allowing the reference dependence of the data--model discrepancy to be assessed.

\bmhead{Acknowledgements}

This work was supported by National Natural Science Foundation of China with Grant No. 12305145 and No. 12547101, Chongqing Natural Science Foundation with Grant No. CSTB2025NSCQ-GPX0745, and Fundamental Research Funds for the Central Universities with Grant No. 2026CDJZKPT-10. R. Ma is supported by the U.S. DOE Office of Science under contract Nos. DE-SC0012704, DE-FG02-10ER41666 and DE-AC02-98CH10886. F. Si is supported by the Alexander von Humboldt Foundation through the Humboldt Research Fellowship for Postdocs with Ref 3.5 - CHN - 1246541 - HFST-P.

\bibliography{references}

\end{document}